\documentclass[namedreferences]{SolarPhysics}
\usepackage[optionalrh,solaenum]{spr-sola-addons} % For Solar Physics 
\usepackage{graphicx}        % For eps figures, newer & more powerfull
\usepackage{color}           % Fo color text: \color command
\usepackage{url}            % For breaking URLs easily trough lines
\def\arcdeg{\hbox{$^\circ$}}
\def\arcsec{\hbox{$^{\prime\prime}$}}
\def\gtrsim{\mathrel{\hbox{\rlap{\hbox{\lower4pt\hbox{$\sim$}}}\hbox{$>$}}}}
\def\sun{\hbox{$\odot$}}
\newcommand{\aap}{    {\it Astron. Astrophys.}}
\newcommand{\aaps}{   {\it Astron. Astrophys. Suppl.}}

\newcommand{\apj}{    {\it Astrophys. J.}}
\newcommand{\apjl}{    {\it Astrophys. J. Lett.}}

\newcommand{\solphys}{{\it Solar Phys.}}

\newcommand{\ssr}{    {\it Space Sci. Rev.}}

\newcommand{\apjs}{    {\it Astrophys. J. Suppl.}}

\begin{document}

\begin{article}

\begin{opening}

\title{Soft X-ray Fluxes of Major Flares Far Behind the Limb as 
Estimated Using STEREO EUV Images }

\author{N.V.~\surname{Nitta}$^{1}$\sep
        M.J.~\surname{Aschwanden}$^{1}$\sep
        P.F.~\surname{Boerner}$^{1}$\sep
        S.L.~\surname{Freeland}$^{1}$\sep
        J.R.~\surname{Lemen}$^{1}$\sep
        J.-P.~\surname{Wuelser}$^{1}$}
\runningauthor{N.V. Nitta}
\runningtitle{Soft X-ray Fluxes of Flares Far Behind the Limb}

   \institute{$^{1}$ Lockheed Martin Solar and Astrophysics
     Laboratory, A021S, Bldg 252, 3251 Hanover Street, Palo Alto, CA
     94304 USA
                     email: \url{nitta@lmsal.com,
                       aschwanden@lmsal.com, boerner@lmsal.com,
                       freeland@lmsal.com, lemen@lmsal.com, wuelser@lmsal.com} \\
             }

\begin{abstract}

With increasing solar activity since 2010, many flares from the
backside of the Sun have been observed by the {\it Extreme Ultraviolet
  Imager} (EUVI) on either of the twin STEREO spacecraft.  Our
objective is to estimate their X-ray peak fluxes from EUVI data by
finding a relation of the EUVI with GOES X-ray fluxes.  Because of the
presence of the Fe~{\sc xxiv} line at 192~\AA, the response of the
EUVI 195~\AA\ channel has a secondary broad peak around 15~MK, and its
fluxes closely trace X-ray fluxes during the rise phase of flares.  If
the flare plasma is isothermal, the EUVI flux should be directly
proportional to the GOES flux.  In reality, the multithermal nature of
the flare and other factors complicate the estimation of the X-ray
fluxes from EUVI observations.  We discuss the uncertainties, by
comparing GOES fluxes with the high cadence EUV data from the {\it
  Atmospheric Imaging Assembly} (AIA) on board the {\it Solar Dynamics
  Observatory} (SDO).  We conclude that the EUVI 195~\AA\ data can
provide estimates of the X-ray peak fluxes of intense flares ({\it e.g.},
above M4 in the GOES scale) with uncertainties of a factor of a few.
Lastly we show examples of intense flares from regions far behind the
limb, some of which show eruptive signatures in AIA images.

\end{abstract}
\end{opening}
%-------------------------------------------------

\section{Introduction}
     \label{S-Introduction} 
The {\it Solar Terrestrial Relations Observatory}
(STEREO; \citeauthor{Kaiser08}, \citeyear{Kaiser08}) was launched in
October 2006, and continues to deliver unique views of the Sun
not accessible from any Earth-bound instruments.  STEREO
consists of the Ahead (A) and Behind (B) spacecraft, 
which drift about 22$\arcdeg$ a year in opposite directions from 
the Sun-Earth line.  Both are equipped with nearly identical copies of 
the {\it Extreme Ultraviolet Imager} (EUVI; \citeauthor{Wuelser04}, \citeyear{Wuelser04}) 
as part of the {\it Sun Earth Connection Coronal and Heliospheric
Investigation} (SECCHI; \citeauthor{Howard08}, \citeyear{Howard08}).
In combination with images from the {\it Atmospheric Imaging Assembly}
(AIA; \citeauthor{Lemen12}, \citeyear{Lemen12}) on board the {\it Solar Dynamics Observatory} (SDO;
\citeauthor{Pesnell12}, \citeyear{Pesnell12}) 
we have observed the full longitudes of the solar corona in EUV since
February 2011.

With the increasing separation of STEREO from the Sun-Earth line
together with increasing solar activity since 2010, 
the EUVI has observed many flares far behind the limb from Earth view.  
Their X-ray emission is therefore completely occulted by the limb, and one of
the frequently asked questions is ``what would be the magnitude of such a
flare in soft X-rays?''  Data from the GOES {\it X-ray Sensor} 
(XRS) have played a major role in our understanding of the physics
of solar flares as they have been 
widely used to derive plasma parameters 
({\it e.g.}, \citeauthor{Feldman96}, \citeyear{Feldman96};
\citeauthor{Battaglia05}, \citeyear{Battaglia05};
\citeauthor{Ryan12}, \citeyear{Ryan12}).
The flare class defined by the peak flux in the 1\,--\,8~\AA\ band 
is generally thought to be a good measure of the released energy.  
In particular, the so-called X-class
($F_{1-8} \gtrsim 10^{-4}$~W m$^{-2}$) flares are often treated with
special interests ({\it e.g.}, \citeauthor{Sudol05}, \citeyear{Sudol05};
\citeauthor{LaBonte07}, \citeyear{LaBonte07}),
because they release enormous amount of energy, and some of them
may have a higher potential to perturb the heliosphere.

In this article, we explore the possibility to use EUVI data
to estimate the GOES X-ray peak fluxes of flares far behind the limb. 
% As shown in the next section, 
The response of the EUVI
with temperature is vastly different from that of the GOES XRS.  However, during
flares, one of the four channels, namely the one that encompasses the
Fe~{\sc xii} complex at 195~\AA\ ($\approx$1.5~MK), should 
observe emissions from mainly
hot ($>$10~MK) plasma (see, for example, \citeauthor{Dere79}
\shortcite{Dere79} who derived a typical differential emission measure of
flares) due to the Fe~{\sc xxiv} line at 192~\AA, which has the peak
temperature of 15\,--\,20~MK, 
depending on the assumed ionization equilibrium.  Therefore we focus
on this EUVI channel to estimate the X-ray flux.  In the next
section, the temperature response of the EUVI is compared to
that of the AIA.  In Section 3, we use AIA data to study the origin and extent of
the uncertainties that limit the accuracy of our work.  This is
followed in Section 4 by our main objective of finding an empirical relation of the
EUVI with the GOES XRS fluxes.  In Section 5 we show examples of
flares from the backside that are estimated to be intense.   Some of
them leave observable signatures in AIA data.  We summarize the work in
Section 6, which also lists items that may possibly improve this work.

\section{Temperature Response} %%%%%%%%%%%%%%%%%%%%%%%%%%%%%%%%%%%%%%%%
      \label{T_Resp}      

The EUVI observes the Sun in four channels of EUV passbands that are
similar to those of the {\it Extreme-ultraviolet Imaging
Telescope} (EIT; \citeauthor{Boudin95}, \citeyear{Boudin95}) on the 
{\it Solar and Heliospheric
Observatory} (SOHO).  They are 171~\AA\ (Fe~{\sc ix}, Fe~{\sc x}), 
195~\AA\ (Fe~{\sc xii}, Fe {\sc xxiv}),
284~\AA\ (Fe~{\sc xv}) and 304~\AA\ (He~{\sc ii}).  Their
temperature responses are plotted in Figure 1 as solid lines.  
Those for the AIA are plotted as dashed lines.
In order to gain more comprehensive understanding of the
structures and dynamics of the corona and transition region, 
the AIA has three more EUV channels,
namely 94~\AA\ (Fe~{\sc xviii}), 131~\AA\ (Fe~{\sc viii}, 
Fe~{\sc xxi}), and 335~\AA\ (Fe~{\sc xvi}).  Additionally, the AIA replaces
the 284~\AA\ channel used by previous missions with the
211~\AA\ channel that contains the Fe~{\sc xiv} line, but the peak
temperature is similar.  

\begin{figure}    %%%%%%%%%%%%%%%%%% FIGURE 1 
 \centerline{\includegraphics[width=0.99\textwidth,clip=]{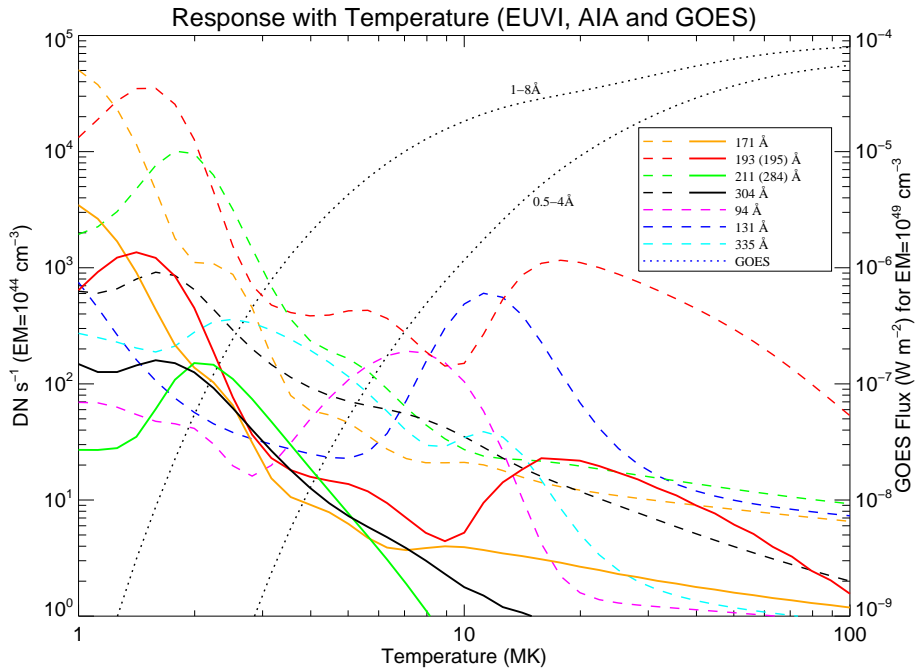}
              }
              \caption{Temperature response of EUVI on STEREO-A (solid lines), AIA
                (dashed lines) and GOES-14 (dotted lines).  They are
                calculated using the CHIANTI atomic database version
                7.0 (Dere {\it et al.}, 1997; Landi {\it et al.}, 2012).
% \cite{Dere97,Landi12}. 
}
   \label{t_resp_euvi_aia_xrs}
\end{figure}

The two bands of the GOES XRS (plotted as dotted lines)
have monotonically increasing
response with temperature.  Although none of these EUV channels has 
a similar temperature response,
some of them have either primary or
secondary peaks at high temperatures that characterize flares 
\cite{Dere79}.  Although AIA's
131~\AA\ channel (peaking at $\approx$10~MK) may be most often
correlated with the XRS as indicated in the next section, it is clear
that the 195~\AA\ is the only EUVI channel that has a distinct peak of
response above 10~MK.  This broad peak around 15~MK is due to the
Fe~{\sc xxiv} line at 192~\AA. Therefore we use this channel exclusively for
estimating the X-ray peak fluxes of flares not observed by GOES.

\begin{figure}    %%%%%%%%%%%%%%%%%% FIGURE 2 
 \centerline{\includegraphics[width=0.99\textwidth,clip=]{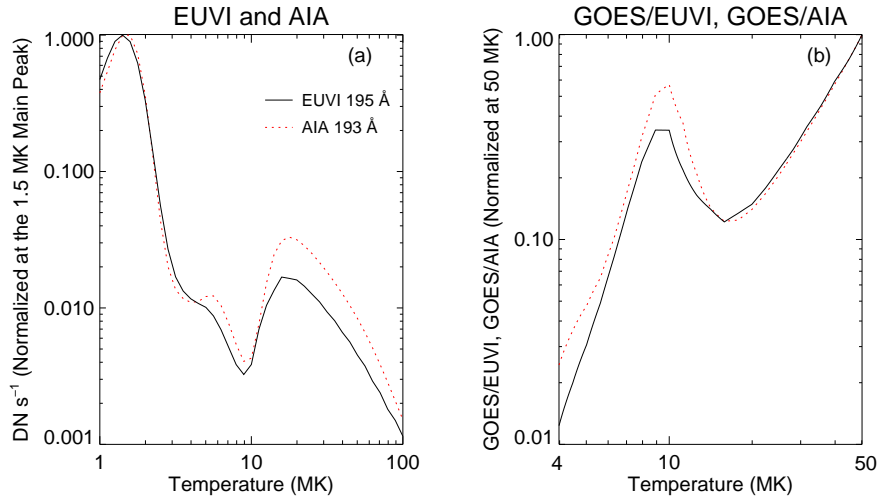}
              }
              \caption{(a) Comparison of EUVI's 195~\AA\ channel with AIA's
              193~\AA\ channel, normalized at the respective primary peaks
              at $\approx$1.5~MK due to Fe~{\sc xii}.  Because AIA's
              passband is shifted to shorter wavelength, the secondary
              peak around 15~MK is more pronounced.  (b) Although 
              the GOES response monotonically increases with
              temperature (Figure~1), the ratio of GOES with AIA and
              EUVI has a dip due to the secondary peak of the latter instruments. }
   \label{t_resp_195}
   \end{figure}

The high temperature response of the
195~\AA\ channel is only secondary to the main peak at $\approx$1.5~MK. 
Figure~2 gives a closer view of the response of EUVI's
195~\AA\ channel in comparison with that of AIA's
193~\AA\ channel.  We expect differences because AIA's passband is
shifted to shorter wavelengths.  As the name indicates, the AIA
channel's effective area peaks at 193~\AA\ rather than at 195~\AA, and
thus has a higher response at the wavelength of 
the Fe~{\sc xxiv} line.  Figure~6 of 
\citeauthor{Howard08} \shortcite{Howard08} and
Figure~8 of \citeauthor{Boerner12} \shortcite{Boerner12} give the
effective areas of the EUVI and AIA, respectively.
Specifically, the  response to 15~MK plasma
relative to the main response to 1.5~MK plasma is about a factor of
two smaller for EUVI than for AIA (Figure~2(a)).  The small difference
of the effective area also affects the flux at a given temperature with
respect to the GOES XRS flux.   
The GOES/EUVI and GOES/AIA ratios
both decrease from 10~MK to 15~MK and increase above 15~MK, but reach
the level of 10~MK at different temperatures as will be shown later in
Figures~5 and 7.
According to Figure~2(b), the GOES/AIA
ratio at 35~MK is still small than that at 10~MK, whereas the
GOES/EUVI ratio at 30~MK is already comparable to that at 10~MK.

\section{AIA-GOES Relation} %%%%%%%%%%%%%%%%%%%%%%%%%%%%%%%%%%%%%%%%
      \label{Data}      
 
In order to understand various uncertainties that may sometimes be
critical in our objective of estimating the GOES X-ray peak flux from
EUVI 195~\AA\ data, we first compare light curves of flares from the
AIA with
those from the GOES XRS.  The AIA takes images every 12 seconds in all the EUV
channels, whereas the typical cadence of EUVI 195~\AA\ images is
5 minutes.  We select all flares above the GOES C3
($F_{1-8}$ = 3$\times$10$^{-6}$~W m$^{-2}$) level between June 2010 and
September 2012.  In order to be free of the uncertainty from 
occultation by the limb, we limit to only those flares whose 2-D radial
distance is less than 0.997$r_{\sun}$.  We used the 
the Heliophysics Event Registry \cite{Hurlburt12} for the flare locations.

\begin{figure}    %%%%%%%%%%%%%%%%%% FIGURE 3
 \centerline{\includegraphics[width=0.99\textwidth,clip=]{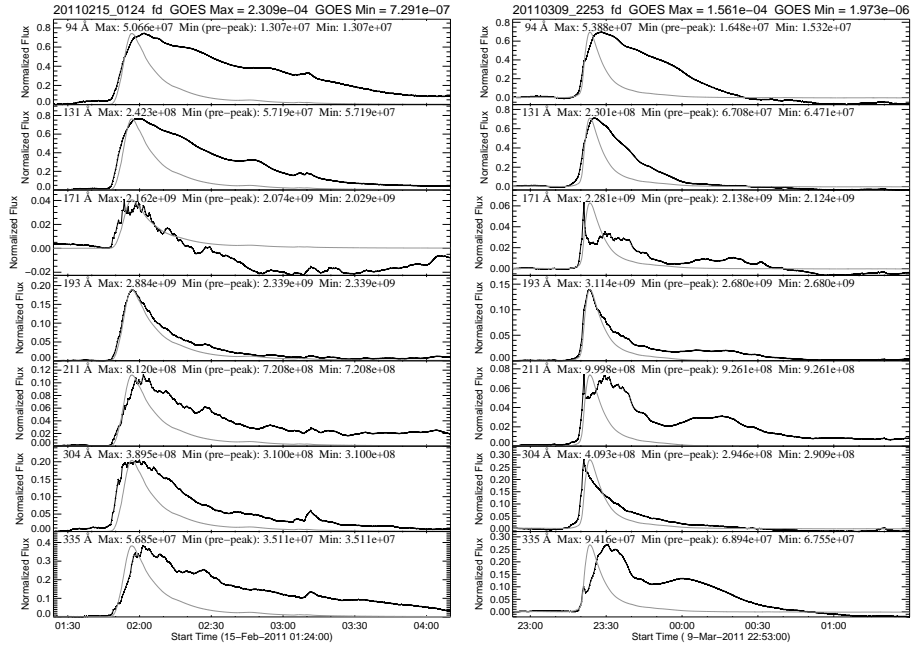}
              }
              \caption{Time variations of fluxes in AIA images. The
                channels plotted are (from top to bottom): 94~\AA,
                131~\AA, 171~\AA, 193~\AA, 211~\AA, 304~\AA, and
                335~\AA.  The
                preflare minimum is subtracted from each time profile.
               The GOES~1\,--\,8~\AA\ light curves is overplotted in
               gray, scaled to [0,maximum] of each EUV light curve. }
   \label{two_flares}
   \end{figure}

Light curves are produced from the headers of individual level-1 full-disk
images, which contain the average pixel values.  
To isolate the emission from the flare,
we subtract the minimum flux during the
30 minute period preceding the GOES peak time. In Figure~3, we show 
two examples of the light curves.  They are X-class flares
(SOL2011-02-15T01:56 and SOL2011-03-09T23:23)\footnotemark.  The GOES XRS light
curves are also shown normalized to the positive scale of the EUV
flux.  In both examples, time variations of 
the 193~\AA\ flux (in the fourth row) is found to be well correlated
with those of
the X-ray flux up
to the GOES peak.  After the peak, the 193~\AA\ flux does not 
decay as fast as the GOES flux presumably because of contributions from cooler
plasma.  
The first flare was
quite eruptive and associated with an energetic coronal mass ejection (CME)
({\it e.g.}, \citeauthor{Schrijver11}, \citeyear{Schrijver11}), whereas the
second one was confined without a CME.  Whether the flare is
associated with a CME may affect our purpose because of
the associated coronal dimming that is pronounced at temperatures of 1\,--\,2~MK.
Figure~3 indicates that the effect of dimming in the 193~\AA\ channel 
may not be substantial, at least for intense flares.

\footnotetext{For SOL identification convention, see {\it Solar Phys.}
  {\bf 263}, pp.1\,--\,2, 2010.}

Concerning other channels, the EUV peak is delayed at 94~\AA\ and
335~\AA\ with respect to the
GOES probably because plasma that cools from $>$10~MK to the
respective response peaks (see Figure~1) also contributes to the peaks
in the light curves.  At low temperatures, most notably at
304~\AA\ but also at 171~\AA\ and even at 211~\AA, the EUV peak 
precedes the XRS peak, reflecting the impulsive phase in which
non-thermal electrons collisionally thermalize plasma at loop
foot-points ({\it e.g.}, \citeauthor{Fletcher11}, \citeyear{Fletcher11}).
The dimming is most pronounced at 171~\AA, but its relation with the CME
is not clear.  In the EUV wavelengths, the flux tends to decay much
more slowly than in soft X-rays, and there are even secondary peaks.
These tendencies seem to be least prominent in the 193~\AA\ channel
as far as the examples in Figure~3 are concerned.

\begin{figure}    %%%%%%%%%%%%%%%%%% FIGURE 4
 \centerline{\includegraphics[width=0.99\textwidth,clip=]{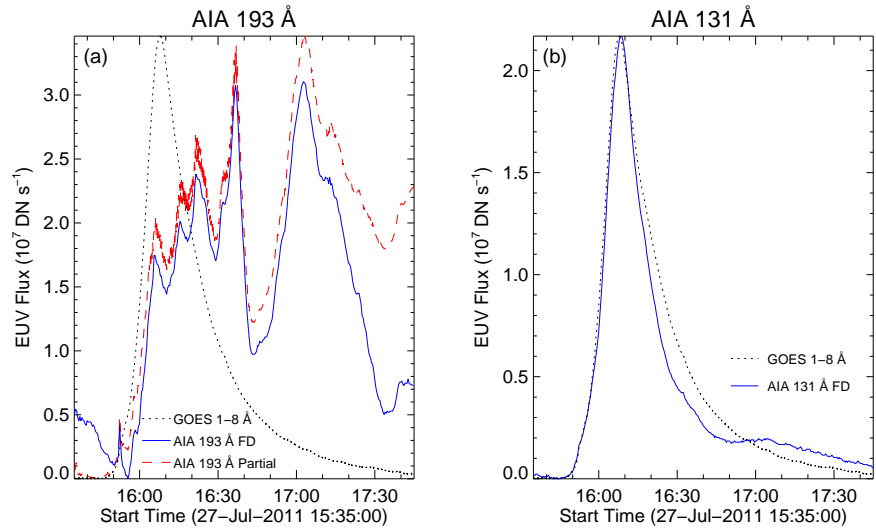}
              }
              \caption{Light curves of AIA images in (a) 193~\AA\ and (b)
                131~\AA\ channels for a M1.1 flare in comparison with
                the GOES soft X-ray time variations.  In 193~\AA,
                there are variations that do not track the GOES,
                although a good match is seen in 131~\AA.  Even if the
                field of view is restricted to a sub-field around the
                flare (see the line in red) as opposed to full-disk
                (FD), the variations observed
                in 193~\AA\ remain, suggestive of increase
                of low-temperature material in the flare that is not
                observed by GOES.    }
   \label{lc_193_131}
   \end{figure}

For less intense flares, there are more complications in the
193~\AA\ data due primarily to the flux of cool plasma that 
contributes to the main
response around 1.5~MK.  In full-disk
193~\AA\ images, the flare contribution is typically only 
$\approx$10\% of the total flux in a X1 flare, meaning that it can be
only 0.5\% for a C5 flare.  Therefore background subtraction produces
significant uncertainties, especially for less intense events.  
Variations of coronal plasma outside the flare region may be excluded by analyzing
the fluxes in a small area surrounding the flare.  However, this does
not always work for flares less intense than, for example, the GOES M3\,--\,M4
level
($F_{1-8}$ = (3\,--\,4)$\times$10$^{-5}$~W m$^{-2}$).  
Figure~4(a) shows that the light curves are not well
correlated between the GOES and 193~\AA\, and that this does not
change substantially if we use the flux within a sub-field
(300$\arcsec \times$300$\arcsec$).  
Furthermore, we could not find a sub-field in which
the first peak that corresponds to the GOES peak is higher than the
second peak around 17:03~UT. 
This example thus indicates that
some flares can have a higher emission measure at 1.5~MK relative to 
20~MK than found by \citeauthor{Dere79} \shortcite{Dere79}.
% elavated emission measure at low temperatures around
% 1.5~MK.
% ({\it e.g.},Dere {\it et al.} 1979). 
For this M1.1 flare
(SOL2011-07-27T16:07), the 131~\AA\ flux is still well-correlated with
the GOES (Figure~4(b)).  One of the reasons that smaller flares
correlate better with the 131~\AA\ than with the 193~\AA\ may be that
they are cooler, and thus reach temperatures of Fe~{\sc xxi} but not 
those of Fe~{\sc xxiv}.

\begin{figure}    %%%%%%%%%%%%%%%%%% FIGURE 5
 \centerline{\includegraphics[width=0.99\textwidth,clip=]
{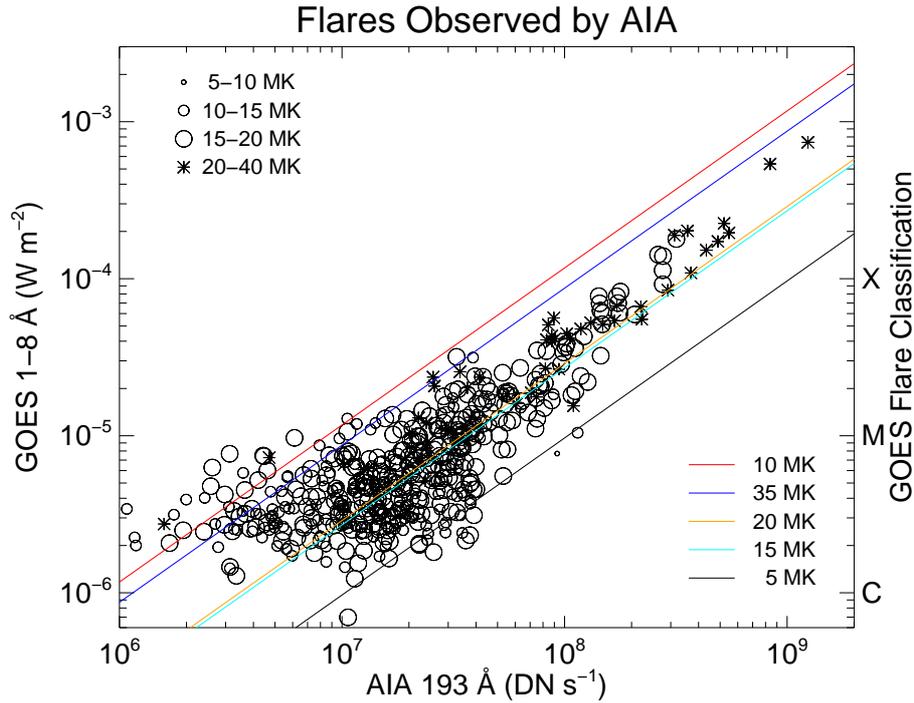}
              }
              \caption{Scatter plot of the GOES 1\,--\,8~\AA\ and AIA
                193~\AA\ background-subtracted fluxes of 540 flares.  In each flare, the GOES-AIA
                pair is made around the X-ray peak time.  The five
                lines show how the relation of the GOES and AIA fluxes
              depends on the temperature of the flare.  The order
              reflects the AIA response relative to the GOES response
              (see Figure~2(b)). The size of
              the symbols represents the flare temperature as
              determined by the GOES 0.5\,--\,4~\AA\ and
              1\,--\,8~\AA\ ratio. }
   \label{aia_xrs}
   \end{figure}

Coronal dimming can also play a major role in
less intense flares than those shown in Figure~3.  Out of the 609
flares above C3 and observed in AIA's normal mode, 69 flares, mostly
C-class, show less 193~\AA\ flux at the GOES peak than during the preflare
interval.  The most intense flare in this categoy is M2.5.  
The frequency of a smaller flux at the flare peak is 
not not much different if we include
only a small area around the flare because of deeper dimming closer
to the flare.  

Cooling plasma from previous flares can contribute to the main
response of the 193~\AA\ channel, making it difficult to determine the
background level.  This is especially true when the increase of flux
due to the flare is only a few percent of less.  Therefore, flares
that occur one after another can pose a serious problem for comparing
EUV and X-ray fluxes.  

In Figure~5, we show a scatter plot of the AIA 193~\AA\ and GOES XRS
1\,--\,8~\AA\ fluxes around the GOES peak times.  Both fluxes reflect
pre-flare background subtraction.  This includes all the 540
flares that have positive 193~\AA\ fluxes. The time difference between
the AIA and GOES fluxes is only up to the three-second sampling rate
of the latter.  Different symbols are used to indicate 
the GOES XRS channel ratio temperature \cite{White05}.
Pearson's correlation coefficient for the fluxes in the logarithmic
scales is 0.81 for the entire population and 0.92 for the 51 flares above M3. 
The lines represent the expected relations between the GOES and AIA
fluxes for flares characterized by single temperatures (see
Figure~2(b)).  Most points above the M3 level fall between $T=$15~MK
and $T=$35 MK lines.  The GOES channel ratio temperature is generally
above 15~MK for these flares.  The scatter in this plot could be made
smaller first by limiting the area and then by conducting
pixel-to-pixel differential measure analyses and extracting only
those pixels that contain hot plasma to be observed by GOES.  It is
beyond the scope of this article, although such an effort could be
instructive.

\section{GOES EUVI Relation} %%%%%%%%%%%%%%%%%%%%%%%%%%%%%%%%%%%%%%%%
      \label{GOES-EUVI}

\begin{figure}    %%%%%%%%%%%%%%%%%% FIGURE 6
 \centerline{\includegraphics[width=0.99\textwidth,clip=]{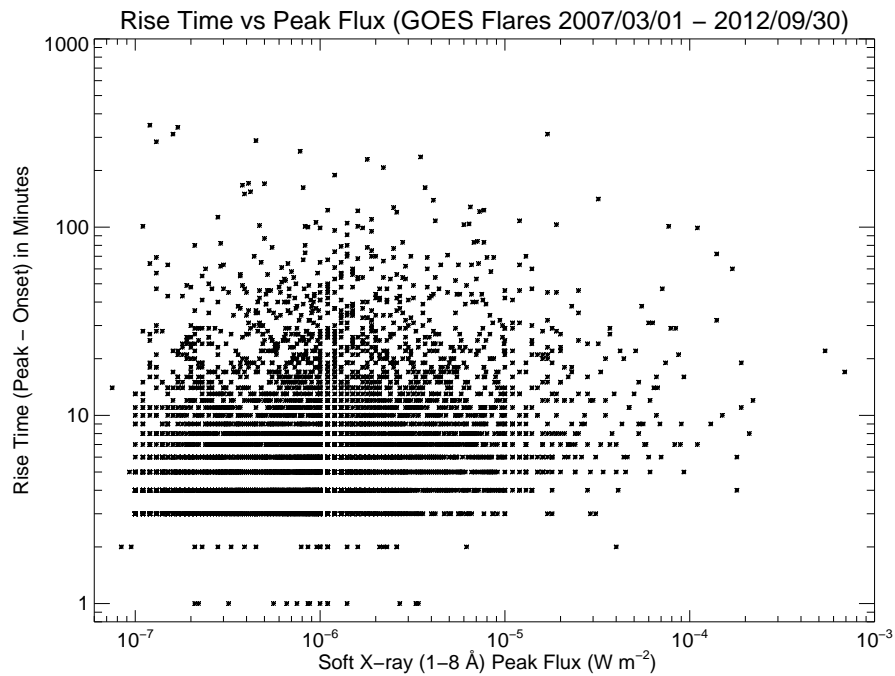}
              }
              \caption{The distribution of the flare rise time from
                the NOAA flare list. }
   \label{rise_time}
   \end{figure}

Now we apply the same analysis on EUVI 195~\AA\ data. 
The most significant difference of these images from AIA images is that
they are taken much less frequently, typically once every five
minutes.  This may mean a poorer correlation with the GOES XRS flux
because the next image after the flare onset can
be after the flare peak, by which time the excess in EUV flux may
already start.  In Figure~6, for all the flares included in the NOAA
flare list, we plot the time difference between the soft X-ray peak
and onset as a function of the X-ray peak flux.  Most flares above the M4
level have the rise time longer than 4~minutes, so it is likely that
for most intense flares the rising phase is included 
where we expect a good correlation between GOES
and EUVI 195~\AA.  There were intense flares in 2007 up to M8
\cite{Aschwanden09, Nitta13}, but we cannot include them because their
rise times are typically less than ten minutes, and 195~\AA\ images
were taken only once every ten minutes.  Until August 2009, EUVI's ``primary
wavelength'' was 171~\AA.

\begin{figure}    %%%%%%%%%%%%%%%%%% FIGURE 7
 \centerline{\includegraphics[width=0.99\textwidth,clip=]
{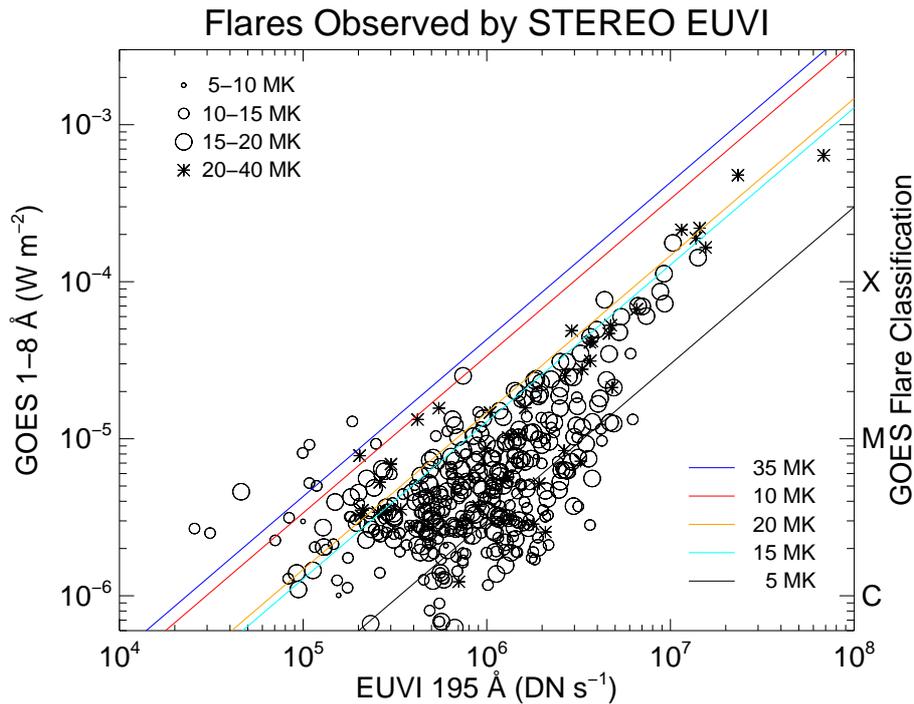}
              }
              \caption{Scatter plot of the GOES 1\,--\,8~\AA\ and EUVI
                195~\AA\ fluxes of $\approx$450 flares.  Only those
                flares observed by AIA and not occulted from STEREO's
                view are plotted.  The same format as Figure~5.  The
                order of the lines representing five temperatures
                reflects the EUVI response relative to the GOES
                response (see Figure~2(b)).}
   \label{euvi_xrs}
   \end{figure}

From the same set of flares that are plotted in Figure~5, we select only
those that were observed on disk by one or both 
STEREO spacecraft.  This is achieved within SolarSoft \cite{Freeland98} 
by combining the flare locations
on flares that come from the Heliophysics Event Registry
\cite{Hurlburt12} with the STEREO orbital information.  
We also drop those flares
in which the time difference of the EUVI image from 
the GOES peak time is more than
five minutes.  This leaves a total of 450 flares with Ahead and Behind combined.  
No distinction is made between the EUVI flux from Ahead and Behind, 
since the response of the 195~\AA\ channel to 10~MK differs by 
only $\approx$4\%.  For the flares that
were observed without occultation by EUVI on both spacecraft, we
usually use data from Ahead.  We correct the EUVI flux as if the
STEREO were located at the same distance as the Earth from the Sun.

Figure~7 is a plot of the observed GOES and full-disk EUVI fluxes.  
% We find a correlation that is nearly as tight as that between the
% GOES XRS and AIA fluxes, even though the EUVI's response to 15~MK
% relative to that to 1.5~MK is twice as small as in the case of the
% AIA (see Figure~2(a)). 
Pearson's correlation coefficient is 0.67 for the entire population,
but as high as 0.94 for the 32 flares above M3.
One unresolved issue, however, is that most points representing intense flares
lie around or below the $T=$15~MK line ({\it e.g.}, the rightmost point, 
which is the X6.9 flare of SOL2011-08-09T08:05 \cite{Asai12}), suggesting the need for fine
tuning of the EUVI absolute flux calibration.

\begin{figure}    %%%%%%%%%%%%%%%%%% FIGURE 8
 \centerline{\includegraphics[width=0.99\textwidth,clip=]
{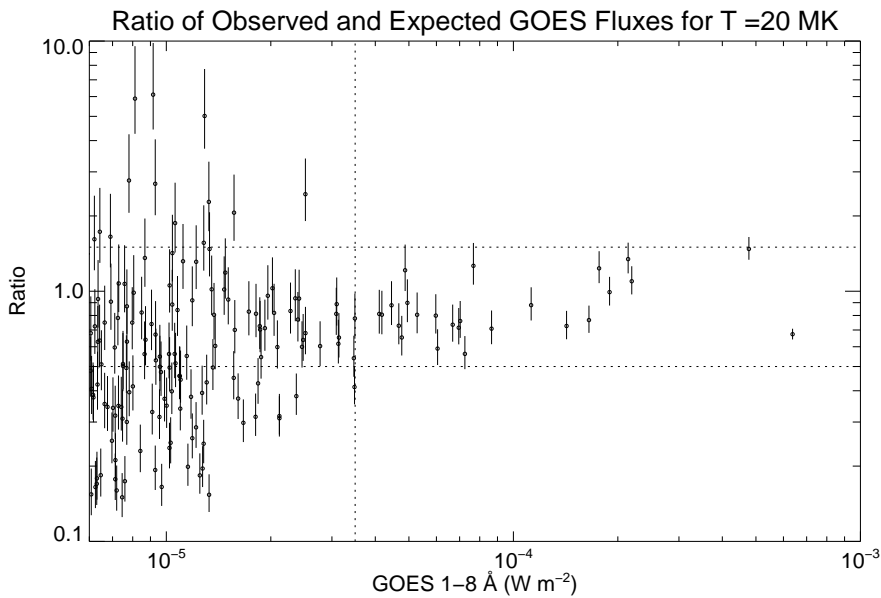}
              }
              \caption{The ratio of the observed GOES flux with the
                expected flux from EUVI 195~\AA\ flux assuming an
                isothermal plasma of 20~MK.}
   \label{euvi_exp_scatter}
   \end{figure}

Irrespective of the absolute calibration, we can now use the points of
intense flares to estimate the GOES peak X-ray flux.  In Figure~8 we show
how much the scatter is from what is expected by assuming that flares have
a single temperature of 20~MK.  Here the error bars reflect a crude
estimate of the uncertainty of the EUVI flux, where it increases from
3\% at $F_{\rm{EUVI(195)}}=10^{8}$ to 50\% at  $F_{\rm{EUVI(195)}}=10^{4}$.
This is based on an inspection of temporal variations of the EUVI flux in
comparison with the XRS flux for all the events included here.

Figure~8 suggests that we can determine the GOES peak
X-ray flux for flares more intense than M4 to within the factor of
three.  The best fit we find is   
\begin{equation}
F_{\rm{GOES}} = 1.39 \cdot 10^{-11} F_{\rm{EUVI(195})}
\end{equation}

The uncertainty is larger for less intense flares as is clear from the
larger scatter.
Nevertheless, it is not much larger than an order of
magnitude, so for convenience we will put the range of 
0.5\,--\,1.5 of Equation (1) for flares with
$F_{\rm{EUVI(195)}} > 4 \cdot 10^{6}$~DN s$^{-1}$ and 0.3\,--\,3 for
flares with $F_{\rm{EUVI(195)}} \lesssim 4 \cdot 10^{6}$~DN s$^{-1}$.

\section{Major Events on the Farside} %%%%%%%%%%%%%%%%%%%%%%%%%%%%%%%%%%%%%%%%
      \label{Examples} 

We have surveyed the mission-long EUVI archive and found a number of
candidates of X-class flares.  Table~1 lists them with their locations
and estimated ranges of the GOES peak X-ray flux.  It also includes a
few less intense flares that have been discussed largely in the
context of multi-spacecraft observations of solar energetic particle
(SEP) events \cite{Dresing12, Rouillard12, Mewaldt12}.  It turns out
that these events (17 January 2010, 21 March 2011, and 3 November 2011) are not
particularly intense in the GOES scale.  But at least the first two
events stand out in terms of one of the attributes of the CME, namely
an EUV wave \cite{Veronig10, Rouillard12}.

We have a few notes concerning the list in Table~1. First, the frequency of
candidates of X-class flares behind the limb mimics that of the
X-class flares actually observed on the visible side.  Single active regions may produce
multiple X-class flares over a wide range of longitude.  
Although we had to wait until 15 February
2011 for the first X-class flare in solar
cycle 24, the flare on 31 August 2010 was probably the first one.

\begin{table}
\caption{List of major events since 2010 that come from regions
  $>$30$\arcdeg$ behind the limb.  These are either
  candidates of X-class flares or multi-spacecraft SEP events. The
  four events that are shown in Figure~9 are marked with asterisks.
}
\label{T-simple}
\begin{tabular}{lcclccl}     % define the column alignment
                           % l: left, c: center, r: right
  \hline                   % horizontal line
Date & Time & A or B & Location & Est. $F_{\rm{GOES}}$ & Range & Refs \\
2010/01/17 & 03:55:40 & B  & S25 E128 & 6.4$\cdot$10$^{-5}$ &
M3.4\,--\,M9.6 & 1, 2\\
2010/08/31 & 20:55:53 & A  & S22 W146 & 1.7$\cdot$10$^{-4}$ &
M8.4\,--\,X2.5 &  \\
2010/09/01 & 21:50:53 & A  & S22 W162 & 1.1$\cdot$10$^{-4}$ &
M5.4\,--\,X1.6 &  \\
2011/03/21 & 02:10:49 & A  & N20 W128 & 3.1$\cdot$10$^{-5}$ &
M1.3\,--\,X1.3 & 3 \\
2011/06/04 & 07:05:58 & A  & N15 W140 & 1.0$\cdot$10$^{-4}$ &
M5.2\,--\,X1.6 &  \\
2011/06/04 * & 21:50:58 & A  & N17 W148 & 8.1$\cdot$10$^{-4}$ &
X4.0\,--\,X12 &  \\
2011/10/23 & 23:15:44 & A  & N19 W151 & 1.1$\cdot$10$^{-4}$ &
X5.3\,--\,X1.6 &  \\
2011/11/03 * & 22:40:44 & B  & N08 E156 & 9.4$\cdot$10$^{-5}$ &
M4.7\,--\,X1.4 & 4  \\
2012/03/26 & 22:16:02 & B  & N18 E123 & 1.6$\cdot$10$^{-4}$ &
M8.2\,--\,X2.5 &  \\
2012/04/29 & 12:45:53 & B, A  & N12 E163 & 1.7$\cdot$10$^{-4}$ &
M8.3\,--\,X2.5 &  \\
2012/07/23 * & 02:30:56 & A  & S15 W133 & 1.5$\cdot$10$^{-4}$ &
M8.2\,--\,X2.5 &  \\
2012/08/21 & 20:10:54 & B, A  & S22 E158 & 1.4$\cdot$10$^{-4}$ &
M6.8\,--\,X2.0 &  \\
2012/09/11 & 07:55:50 & B, A  & S22 E178 & 2.6$\cdot$10$^{-4}$ &
X1.3\,--\,X3.9 &  \\
2012/09/19 & 11:15:49 & B, A  & S15 E171 & 1.8$\cdot$10$^{-4}$ &
M9.1\,--\,X2.7 &  \\
2012/09/20 * & 15:00:48 & B, A  & S15 E153 & 1.2$\cdot$10$^{-3}$ &
X5.8\,--\,X18 &  \\
2012/09/22 & 03:05:48 & B  & S15 E134 & 1.7$\cdot$10$^{-4}$ &
M8.4\,--\,X2.5 &  \\

  \hline
\end{tabular}

References: 1. Veronig {\it et al.} (2010), 2. Dresing {\it et al.} (2012),
3. Rouillard {\it et al.} (2012), 4. Mewaldt {\it et al.} (2012) 

\end{table}

\begin{figure}    %%%%%%%%%%%%%%%%%% FIGURE 9
 \centerline{\includegraphics[width=0.99\textwidth,clip=]
{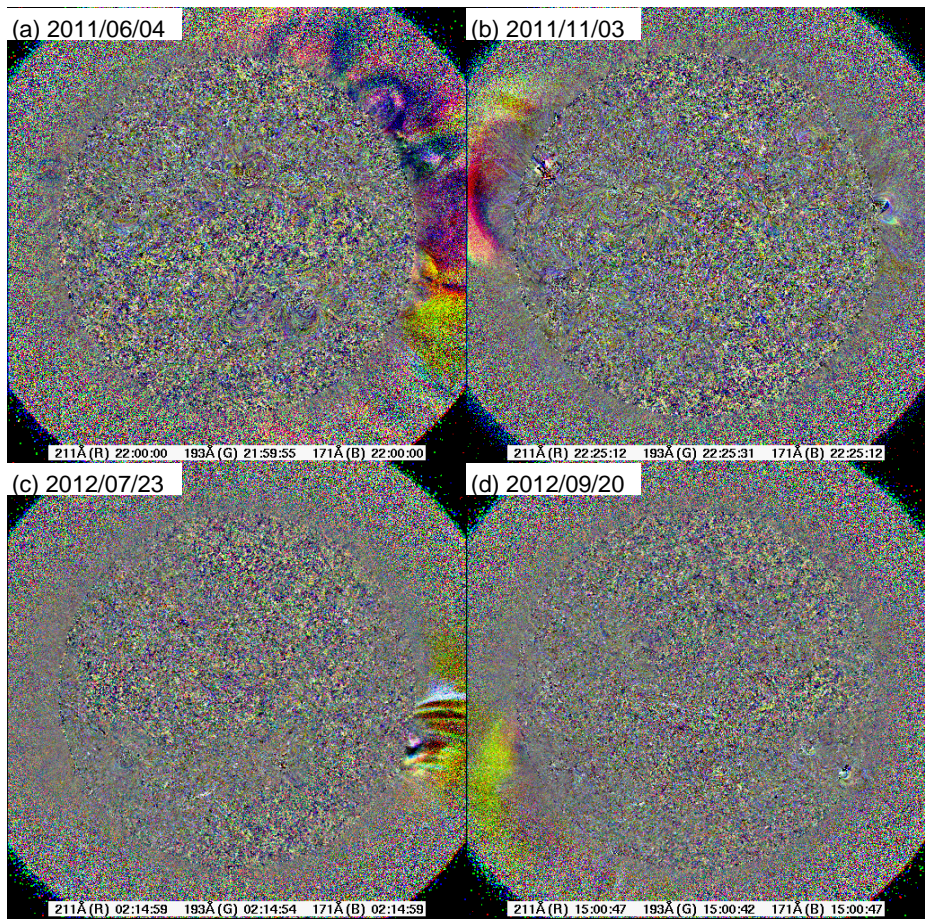}
              }
              \caption{Running ratio images of four events that
                occurred far behind the limb (see Table 1 for their
                locations), and were observed directly by AIA because
                of high cadence and sensitivity.  }
   \label{tri_color}
   \end{figure}

EUVI data of these behind-the-limb flares (from Earth view) show a varied
degree of association with CMEs.  
While some are confined and not associated with a major CME in
white-light coronagraph data, others are highly eruptive, also
accompanying extended dimming.  Note that dimming is usually observed only
after a sharp increase of flux, indicating that the peak
flux is well captured.  Some of the highly eruptive events, despite deep
occultation by the limb, emerge in AIA images either as an eruption itself or in the form
of waves or propagating fronts.  
Figure~9 gives snapshots from the
tri-color running ratio movies that are accessible at 
\url{http://aia.lmsal.com/occulted_events.}  We also point out that the event on
23 July 2012 is more eruptive than that on 20 September 2012,
which is probably the most intense flare in solar cycle 24 as of  the
end of 2012.

\section{Summary and Discussion} %%%%%%%%%%%%%%%%%%%%%%%%%%%%%%%%%%%%%%%%
      \label{S-Discussion} 

We have correlated GOES X-ray and EUV (195~\AA) fluxes of 450 flares as observed by
STEREO, and found an empirical relation, which lets us estimate the
GOES peak X-ray flux of intense events, whose X-ray emission is
occulted by the limb.  There are more than a
dozen candidates of X-class flares as shown in Table~1.  
Many of the events included here to obtain the EUVI-XRS flux relation have 
impulsive time profiles, and in such events
it is possible that we may underestimate the soft X-ray flux 
by missing the real peak because of the EUVI five-minute cadence limitation.

Using AIA observations we study the origins and extents of
uncertainties in the estimation of the peak GOES X-ray flux from EUV
data.  The main difficulty lies in the fact that the Fe~{\sc xxiv}
line represents only a small secondary response peak in the
195~\AA\ (193~\AA) channel.  The main response peak not only picks up
cooling or heating plasma from the present and previous flares in the 
1\,--\,2~MK range but also shows decrease because of coronal dimming
when the flare is associated with a CME.
It appears that the effect of coronal dimming is not substantial until
the soft X-ray peak in flares above M3 or so, but it can be 
significant in less intense flares.  
In extreme cases, the 193~\AA\ flux is found to decrease
in the rising phase of the flare.  The primary response of AIA's 193~\AA\ and EUVI's
195~\AA\ channels is around 1.5~MK, and in order to correlate the
fluxes in these channels with the GOES flux, we should in principle
extract those pixels that contain $\gtrsim$10~MK plasma, by conducting differential
emission measure analysis.  One caveat is that one needs to monitor
full-disk images for sympathetic flaring at a remote region, which may
also contribute to the spatially-integrated GOES flux.

Other related issues include the absolute calibration of the involved
instruments, including the GOES XRS.  We used calibration data for the
XRS on GOES-14, but most flares analyzed in this work were observed by
the XRS on GOES-15, whose response data have not been available
(S. White, personal communication, 2012)\footnotemark.  Also needed is a fine tuning
as to the absolute calibration of EUVI and AIA.  Apart from the 
calibration of the instruments, we have
not considered the possible effect of non-thermal flare emission
(especially on GOES) and
ionization equilibrium of hot plasma.

\footnotetext{The response of the XRS on GOES-15 became available in
  March 2013, around the time of the second referee report, but the
  difference from that of the XRS on GOES-14 seems to be too small to
  affect this work in a significant way.}

% Despite the stated uncertainties, this work produced a way to
% isolate and estimate the peak fluxes of intense ({\it e.g.}, X-class) 
% flares using STEREO EUVI data.  
Lastly, we may ask how significant the GOES X-ray flare classification is for the
interpretation of energetic events.  We have conventionally
and conveniently used the GOES X-ray class as if it represents the
magnitude of the flare.  However, it is argued that the energy
radiated in soft X-rays is only a small fraction of the total
energy released in flares 
({\it e.g.}, \citeauthor{Kretzschmar10}, \citeyear{Kretzschmar10}).
For the total energy budget, we should include presumably 
larger energy that goes into Sun's lower atmosphere.  The fact that even C-class flares
produce white-light emission ({\it e.g.}, \citeauthor{Matthews03},
\citeyear{Matthews03}) indicates that the X-ray flux may not be
correlated with the total energy.  Here we consider the GOES X-ray
flux to be still an important reference, however,
thanks partly to its existence for more than three solar cycles.  As
demonstrated by the recent work of \citeauthor{Ryan12} \shortcite{Ryan12}, the GOES XRS
measurement is useful for collectively deriving flare plasma parameters,
even though it is also possible that we may find in SDO data a
better parameter more intimately correlated with
the total energy that includes energy flow both into the heliosphere
and lower atmosphere.

% we may need to
% include longer wavelength such aoptical continuum) to calculate the
% total energy of flares ({\it e.g.}, .  
% The X-ray flux may be correlated with the
% total released energy, but we know that white-light flares
% from the lower atmosphere could actually be C-class flares in X-rays
% .  For
% practical space weather applications, intense flares are neither
% necessary nor sufficient conditions for energetic CMEs that perturb
% the heliosphere and accelerate high-energy particles detected
% in-situ.  Even though this work was motivated by SEP events observed
% at multiple spacecraft widely separate in longitude, we may need a
% better parameter that is more closely linked with eruptive events than
% the GOES X-ray flux.  We may be able to find it in SDO data.

\begin{acks}
 This work has been supported by the NASA STEREO mission under NRL
 Contract No. N00173-02-C-2035, and by the AIA contract NNG04EA00C
to LMSAL.  We are greateful to the referee for valuable comments that
clarified the motivations of this work.

\end{acks}

\end{article} 

\end{document}